\documentclass[11pt]{article}
\usepackage[a4paper,margin=0.90in]{geometry}
\usepackage[T1]{fontenc}
\usepackage[utf8]{inputenc}
\usepackage{lmodern}
\usepackage{amsmath,amssymb,bm,amsfonts}
\usepackage{authblk}
\usepackage{booktabs,tabularx,array,longtable}
\usepackage{float}
\usepackage{enumitem,placeins}
\usepackage{xcolor}
\usepackage{graphicx}
\usepackage{tikz,caption,cite}
\usepackage{fontawesome5}
\usepackage[most]{tcolorbox}
\usetikzlibrary{arrows.meta,positioning,fit,backgrounds,calc}
\usepackage[colorlinks=true,linkcolor=blue!60!black,
  urlcolor=blue!60!black,citecolor=blue!60!black]{hyperref}

\definecolor{agentblue}{HTML}{DCEFFD}
\definecolor{toolgreen}{HTML}{DFF2E1}
\definecolor{storegold}{HTML}{FFF1C9}
\definecolor{guardred}{HTML}{F9DDDD}

\definecolor{blue_munsell}{rgb}{0.0, 0.5, 0.69}
\usepackage{xspace}

\renewcommand{\r}{\textcolor{blue}{\zeta}}
\definecolor{agentblue}{RGB}{205,225,245}
\definecolor{toolgreen}{RGB}{210,240,215}
\definecolor{guardred}{RGB}{245,215,215}

\usepackage[most]{tcolorbox}
\usepackage{xcolor}
\usetikzlibrary{positioning,calc,fit,backgrounds,arrows.meta,backgrounds}

\title{\textbf{A Catalog of Lepton Mass Textures\\
from Koide and Alikhanov Relations}}
\author{Shaikh Saad}
\affil{Jo\v{z}ef Stefan Institute, Jamova 39, Ljubljana, Slovenia}
\affil{\texttt{\color{blue_munsell}shaikh.saad@ijs.si}}
\date{}

\begin{document}
\maketitle

\begin{abstract}
The empirical Koide relation for charged-lepton masses and the
recently proposed Alikhanov relation for neutrino mass-squared
differences offer intriguing hints toward understanding the
structure of lepton mass matrices. Motivated by these observations, we systematically explore charged-lepton and neutrino mass-matrix textures that satisfy
these relations and confront their predictions for lepton mixing
with experimental data.
For the charged-lepton sector, we investigate representative
positive Hermitian square-root matrices classified by the
number of common eigenvector directions of their generators:
three, one, or none.
For the neutrino sector, an exhaustive enumeration of real
symmetric integer matrices with entries between $-5$ and $5$
identifies six signed-permutation classes with $m_1=0$ and
six with $m_1>0$ that satisfy the Alikhanov relation. Combining these neutrino textures with the charged-lepton
constructions, allowing for different relative flavor
orientations and phase alignments, yields $400$ viable
structural pairs consistent with the measured lepton mixing
parameters. All retained solutions satisfy the adopted cosmological bound
on the sum of neutrino masses and existing laboratory constraints.
In particular, the full-rank neutrino textures predict
$m_1\simeq16.77$--$17.22\,\mathrm{meV}$.
Representative viable solutions yield effective Majorana masses
$m_{\beta\beta}\simeq1.29$--$4.03\,\mathrm{meV}$ and
$4.07$--$18.67\,\mathrm{meV}$ for the $m_1=0$ and $m_1>0$
branches, respectively, providing
potential targets for future high-sensitivity searches.
\end{abstract}

{\small\tableofcontents}
\section{Introduction}
\label{sec:introduction}

The pronounced hierarchical structures of charged-fermion masses and quark mixing angles remain among the longstanding mysteries of particle physics. The Standard Model (SM) offers no explanation for these patterns; instead, the fermion masses and mixing parameters are treated as free inputs of the theory. The neutrino sector, on the other hand, exhibits a rather different pattern, with comparatively large mixing angles and a distinct hierarchy of mass-squared differences. Although flavor symmetries beyond the SM may provide an explanation for these intriguing structures, no compelling experimental evidence has yet established the underlying symmetry responsible for them.

Nevertheless, certain empirical relations among lepton masses have attracted considerable attention. In particular, Koide~\cite{Koide:1983qe,Foot:1994yn,Koide:2000zi,Koide:2005ep,Koide:2006dn,Koide:2007eu,Haba:2007vy,Xing:2006vk,Sumino:2008hy,Koide:2012kn} identified a remarkable relation among the experimentally measured charged-lepton pole masses, given by
\begin{equation}
\boxed{
\frac{m_e+m_\mu+m_\tau}
{\left(\sqrt{m_e}+\sqrt{m_\mu}+\sqrt{m_\tau}\right)^2}
=\frac{2}{3}.
}
\label{eq:koide}
\end{equation}
More recently, Alikhanov~\cite{Alikhanov:2026jmw,Alikhanov:2026suu,Alikhanov:2026bhi} pointed out an intriguing relation between the two experimentally measured neutrino mass-squared differences,
\begin{equation}
\boxed{
\frac{\sqrt{\Delta m_{31}^2}+\sqrt{\Delta m_{21}^2}}
{\sqrt{\Delta m_{31}^2}-\sqrt{\Delta m_{21}^2}}
=\sqrt{2}
}
\label{eq:alikhanov}
\end{equation}
where normal neutrino mass ordering is assumed. 

Both relations exhibit remarkable numerical agreement with
experimental observations, suggesting that they may provide
valuable insights into the structure of lepton mass matrices.
Motivated by these observations, we systematically explore
charged-lepton and neutrino mass-matrix textures that accommodate
the Koide and Alikhanov relations, investigating their
phenomenological implications and confronting the resulting
predictions for lepton masses and mixing parameters with
current experimental data.
Our analysis reveals a remarkably rich landscape of
phenomenologically viable lepton mass textures.
Within our bounded search, we identify 402 distinct
prediction families compatible with the observed lepton
mixing parameters, vastly extending the two explicit
constructions presented very recently in Ref.~\cite{Strumia:2026aod}.
Remarkably, the complete set of allowed predictions
can be reproduced using only 400 representative families.

\begin{table}[htbp]
\centering
\begin{tabular}{lll}
\toprule
Quantity & Experimental values$\pm 1\sigma$ interval & $3\sigma$ ranges\\
\midrule
$m_e$ & $\left(0.51099895069\pm 0.00000000016\right)~\mathrm{MeV}$ & -\\
$m_\mu$ & $\left(0.1056583755\pm 0.0000000023\right)~\mathrm{GeV}$ & -\\
$m_\tau$ & $\left(1.77693\pm 0.00009\right)~\mathrm{GeV}$ & -\\
\midrule
$\sin^2\theta_{12}$& $0.3088^{+0.0067}_{-0.0066}$ & $\left(0.2893,\,0.3295\right)$ \\
$\sin^2\theta_{23}$& $0.470^{+0.017}_{-0.014}$ & $\left(0.432,\,0.587\right)$ \\
$\sin^2\theta_{13}$& $0.02249^{+0.00057}_{-0.00057}$ & $\left(0.02070,\,0.02420\right)$\\
\midrule
$\Delta m^2_{21}/10^{-5}$& $7.537^{+0.094}_{-0.10}~\mathrm{eV}^2$ & $\left(7.236,\,7.823\right)~\mathrm{eV}^2$
\\
$\Delta m^2_{31}/10^{-3}$& $2.521^{+0.026}_{-0.018}~\mathrm{eV}^2$ & $\left(2.454,\,2.592\right)~\mathrm{eV}^2$
 \\
\bottomrule
\end{tabular}
\caption{Experimental inputs for charged-lepton masses~\cite{ParticleDataGroup:2026mpi} and neutrino
oscillation parameters~\cite{NUFIT,Esteban:2020cvm}.}
\label{tab:experimental_inputs}
\end{table}

Before proceeding to the next section, we summarize the relevant
experimental inputs in Table~\ref{tab:experimental_inputs}. Throughout this work, we assume a normal hierarchy for neutrino masses. 
Using the central values of the measured quantities, we obtain
\begin{align}
\frac{m_e+m_\mu+m_\tau}
{\left(\sqrt{m_e}+\sqrt{m_\mu}+\sqrt{m_\tau}\right)^2}
&=0.666664,
\qquad \left(\frac{2}{3}\simeq 0.666667\right),
\\
\frac{\sqrt{\Delta m_{31}^2}+\sqrt{\Delta m_{21}^2}}
{\sqrt{\Delta m_{31}^2}-\sqrt{\Delta m_{21}^2}}
&=1.41811,
\qquad \left(\sqrt{2}\simeq 1.41421\right).
\end{align}
Remarkably, both relations exhibit excellent numerical agreement
with the experimental central values. The Koide relation is satisfied
with a relative deviation of approximately $4\times 10^{-6}$,
whereas the Alikhanov relation agrees at the level of about $0.3\%$. Additionally, the Alikhanov relation is also fully consistent with the recent
JUNO measurements based on $207.2$ live days of data~\cite{JUNO:2026tap}.
Further improvements in experimental precision with increasing
JUNO exposure will provide an important test of this relation.
These intriguing observations motivate our investigation of the
underlying lepton mass textures.

\section{A fresh look at Koide's mass relation}
\label{sec:koide}

\subsection{A systematic prescription}

The Koide relation involves square roots of masses, therefore, a natural form of the charged-lepton mass matrix proposed in Ref.~\cite{Koide:2006dn} is
\begin{equation}
M_e=m_*K^2,\qquad m_*>0,
\end{equation}
with $K$ positive Hermitian. Its positive eigenvalues obey
$m_\ell=m_*\lambda_\ell^2$ (here, $\lambda_\ell$ are positive eigenvalues of the matrix $K$), so the exact Koide relation demands the following:
\begin{equation}
\frac{\operatorname{tr}(K^2)}{[\operatorname{tr}K]^2}= \frac{\lambda^2_e+\lambda^2_\mu+\lambda^2_\tau}
{\left(\lambda_e+\lambda_\mu+\lambda_\tau\right)^2} = \frac23.
\end{equation}

In general, the eigenvalues $\lambda_i$ of any Hermitian matrix $K$
can be expressed as~\cite{Koide:2006dn}
\begin{align}
\lambda_1 = \phi-\psi\sin\xi,\qquad
\lambda_2 = \phi-\psi
\sin\left(\xi+\frac{2\pi}{3}\right),\qquad
\lambda_3 = \phi-\psi
\sin\left(\xi+\frac{4\pi}{3}\right),
\end{align}
where
\begin{equation}
\phi=\frac{1}{3}\operatorname{tr}K,\qquad
\psi=\frac{\sqrt{2}}{3}\sqrt{3\operatorname{tr}(K^2)
-\left(\operatorname{tr}K\right)^2}.
\end{equation}
Therefore, any three-flavor mass spectrum can be parametrized
in terms of the overall scale $\phi$ and the two parameters
$\psi/\phi$ and $\xi$, independently of the structure of $K$.
In particular, imposing the condition
\begin{equation}
\frac{\operatorname{tr}(K^2)}{\left(\operatorname{tr}K\right)^2}=\frac{2}{3}  \label{eq:second}
\end{equation}
yields
\begin{equation}
\psi=\sqrt{2}\,\phi, \qquad \phi>0.
\end{equation}

Since the overall normalization of $K$ is irrelevant for
satisfying Eq.~\eqref{eq:second}, we conveniently choose
$\phi=1$, corresponding to $\operatorname{tr}K=3$.
With this normalization, we consider the following
one-parameter family of Hermitian matrices:
\begin{equation}
\boxed{K(\theta)=I+\sqrt3\bigl(\cos\theta\,T+\sin\theta\,S\bigr).} \label{eq:master}
\end{equation}
Here, $I$ is the $3\times 3$ identity matrix and  $T,S$ are some fixed traceless, orthonormal Hermitian matrices. In general, the angle $\theta$, need not coincide with $\xi$. Note that $\{I/\sqrt{3},T,S\}$ forms an orthonormal set under the trace inner product\footnote{
On the real vector space of Hermitian matrices, the Hilbert--Schmidt
inner product is $\langle A,B\rangle=\operatorname{tr}(A^\dagger B)
=\operatorname{tr}(AB)$.
}, where $T$ and $S$ are fixed traceless Hermitian matrices satisfying 
\begin{equation*}
\operatorname{tr}(T^2)=\operatorname{tr}(S^2)=1,\qquad
\operatorname{tr}(TS)=0.
\end{equation*}
For each fixed generator pair \((T_i,S_i)\), the normalized charged-lepton texture \(K_i(\theta)\) is a one-parameter family.

Our construction, Eq.~\eqref{eq:master}, guarantees
\begin{equation}
\operatorname{tr}K=\lambda_1+\lambda_2+\lambda_3=3,\qquad \operatorname{tr}K^2=\lambda_1^2+\lambda_2^2+\lambda_3^2=6, \qquad \frac{\operatorname{tr}(K^2)}{\left(\operatorname{tr}K\right)^2}=\frac{2}{3},
\end{equation}
independently of $\theta$. These are two constraints on three eigenvalues, leaving one continuous
degree of freedom. 
Thus, the Koide relation is satisfied identically,
leaving a single free parameter $\theta$, which we utilize to reproduce the charged-lepton mass hierarchy, while $m_*$
in $M_e=m_*K^2$ fixes the overall scale.

\subsection{Three possible categories}
Our key organizing criterion for extending the charged-lepton search is
\textbf{how many mutually orthogonal eigenvectors remain fixed throughout
the entire texture family}. This criterion concerns the continuous family before fitting the charged
masses, not the eigenvectors of just one fitted matrix.
Here a fixed \emph{direction} means that a normalized vector $x$, independent
of $\theta$, satisfies $K(\theta)x=\lambda(\theta)x$ for every $\theta$.
Vectors differing only by an overall phase represent the same \emph{direction};
the eigenvalue $\lambda(\theta)$ may vary with $\theta$.

In search of such classification, the form of $K$ given in Eq.~\eqref{eq:master}, is particularly useful since, it has three orthonormal generators
\begin{equation}
G_1=\frac{I}{\sqrt3},\qquad G_2=T,\qquad G_3=S.
\end{equation}
and every nonzero vector is an eigenvector of $I$. Therefore, the simplest classification counts the maximum number $\r$ of mutually
orthogonal eigenvectors common to all three fixed generators. The identity generator has every nonzero vector as an eigenvector and
therefore imposes no additional restriction: it suffices to find the
common eigenvectors of $T,S$. These are exactly the directions that
remain eigenvectors of $K(\theta)$ for every shape angle. The only possible
values are $\r=3,1,0$.\footnote{If two fixed orthonormal eigenvectors are
shared by every Hermitian matrix in the family, Hermiticity makes their
one-dimensional orthogonal complement invariant for every parameter
choice. A fixed unit vector in that complement is therefore a third
common eigenvector. Thus the maximal number of common orthogonal
eigenvectors cannot be two.}
These define three common-eigenvector categories, represented here by
the chosen prescriptions $K_1,K_2,K_3$, respectively. Our classification
is summarized in Table~\ref{tab:common-generator-summary}.

\begin{table}[htbp]
\centering
\begin{tabular}{llc}
\toprule
Prescription & Eigenvectors common to $G_1,G_2,G_3$ & $\r$\\
\midrule
$K_1$ & \textcolor{blue}{Three} (the Fourier vectors) & \textcolor{blue}{3}\\\midrule
$K_2$ & \textcolor{blue}{One} direction: $w=u\times v$ & \textcolor{blue}{1}\\\midrule
$K_3$ & \textcolor{blue}{None} & \textcolor{blue}{0}\\
\bottomrule
\end{tabular}
\captionof{table}{Classification of the three chosen prescriptions by eigenvectors common to all three generators. The count refers to the full shape-angle family before fitting charged lepton masses.}\label{tab:common-generator-summary}
\end{table}

In other words, the above classification\footnote{
The classification of matrix pairs according to their
common eigenvectors is closely related to the theory
of Hermitian matrix pencils and their simultaneous
reducibility, which has been extensively investigated
in the mathematical literature~\cite{Shapiro:1982}.
} is equivalent to asking \textbf{How many mutually orthogonal eigenvector directions remain fixed
as $\theta$ varies throughout the family?}  This is summarized in Table~\ref{tab:eigenvector-angle-dependence}. It classifies the eigenvector behavior of families, \textbf{not all possible
textures}. We consider three explicit texture families, one representative of each possible common-eigenvector category. We do not claim that these are the only possible texture families.
\begin{table}[htbp]
\centering
\begin{tabular}{@{}p{0.21\linewidth}p{0.75\linewidth}@{}}
\toprule
Family & Eigenvector dependence on $\theta$\\
\midrule
$K_1$, $\r=3$ & All three directions remain fixed.\\\midrule
$K_2$, $\r=1$ & One direction remains fixed; the other two rotate within its orthogonal plane.\\\midrule
$K_3$, $\r=0$ & No eigenvector direction remains fixed throughout the family.\\
\bottomrule
\end{tabular}
\captionof{table}{Dependence of eigenvector \emph{directions} on the shape angle $\theta$.}\label{tab:eigenvector-angle-dependence}
\end{table}

\subsection{Case studies}
\label{sec:koide-examples}
We now give one explicit generator pair $(T_i,S_i)$ for each of the
three common-eigenvector categories. These choices specify particular
texture families; the classification alone does not uniquely determine
their matrix entries. Table~\ref{tab:generator-angle-table} summarizes
the three generator pairs and their shape-angle ratios; the matrices,
vectors and coefficients used in the table are defined below.
To distinguish exact Koide normalization from
matching to the measured masses, it is useful to write
\begin{equation}
K_i(\theta)=I+\rho\bigl(\cos\theta\,T_i+\sin\theta\,S_i\bigr),
\qquad \rho=\sqrt3\quad\text{for exact Koide}.
\label{eq:examples-radius}
\end{equation}
For fixed $\rho$, the trace identities are
$\operatorname{tr}K_i=3$ and $\operatorname{tr}K_i^2=3+\rho^2$.
The physical Koide ratio follows on the intervals where all eigenvalues
of $K_i$ are positive. Exact Koide, $\rho=\sqrt3$, is compatible with the
quoted charged-lepton mass uncertainties. The algebraic matching
specified below uses a slightly adjusted fixed radius solely to reproduce
all three supplied central values exactly, not to restore experimental
compatibility.

\begin{table}[htbp]
\centering
\renewcommand{\arraystretch}{2.2}
\begin{tabular}{cccc}
\toprule
Family & $T_i$ & $S_i$ & $\tan\theta$\\
\midrule
$K_1$ ($\r=3$) & $\dfrac{P+P^\dagger}{\sqrt6}$
 & $\dfrac{i(P-P^\dagger)}{\sqrt6}$ & $\tan\varphi$\\
\midrule
$K_2$ ($\r=1$) & $\dfrac{3uu^T-I}{\sqrt6}$
 & $\dfrac{uv^T+vu^T}{\sqrt2}$
 & $\dfrac{\gamma}{\sqrt3(1-\alpha)}$\\
\midrule
$K_3$ ($\r=0$) & $\dfrac{D}{\sqrt2}$ & $\dfrac{H}{\sqrt{18}}$
 & $\dfrac{3b}{a}$\\
\bottomrule
\end{tabular}
\caption{Fixed orthonormal generators and shape-angle ratios for the
three chosen prescriptions. The tangent ratios are independent of the
fixed radius $\rho$; the coefficient relations in the text determine
the quadrant.}\label{tab:generator-angle-table}
\end{table}

For these three examples we use the fixed matrices
\begin{equation}
P=\begin{pmatrix}0&1&0\\0&0&1\\1&0&0\end{pmatrix},\qquad
D=\operatorname{diag}(-1,0,1),\qquad
H=\begin{pmatrix}0&1&2\\1&0&2\\2&2&0\end{pmatrix}.
\label{eq:examples-fixed-matrices}
\end{equation}
For $K_2$ we choose the real orthonormal directions
\begin{equation}
u=(0,0,1)^T,\qquad v=(2,-1,0)^T/\sqrt5,\qquad
u^Tu=v^Tv=1,\quad u^Tv=0.
\label{eq:examples-k2-directions}
\end{equation}
All three displayed generator pairs satisfy
$\operatorname{tr}T_i=\operatorname{tr}S_i=0$,
$\operatorname{tr}T_i^2=\operatorname{tr}S_i^2=1$ and
$\operatorname{tr}(T_iS_i)=0$. The tangent ratios are defined when their
denominators are nonzero; the sine and cosine relations below also
cover those exceptional points.

\medskip
\noindent$\bullet$\;\textbf{$K_1$ ($\r=3$):}
The first choice gives the cyclic Hermitian family
\begin{equation}
K_1=I+\frac{\rho}{\sqrt6}
 \bigl(e^{i\varphi}P+e^{-i\varphi}P^\dagger\bigr),
\qquad \varphi=\theta\pmod{2\pi}.
\label{eq:examples-k1}
\end{equation}
For $\rho=\sqrt3$, its off-diagonal magnitude is $1/\sqrt2$.
This reproduces the cyclic structure of Ref.~\cite{Koide:2000zi}.
Since $P^3=I$ and $P^\dagger=P^2$, both generators are polynomials in
$P$ and commute. Their common orthonormal eigenvectors are
\begin{equation}
f_j=\frac{1}{\sqrt3}(1,\omega^j,\omega^{2j})^T,\qquad
\omega=e^{2\pi i/3},\qquad Pf_j=\omega^j f_j,
\quad j=0,1,2.
\label{eq:examples-fourier-vectors}
\end{equation}
Consequently,
\begin{equation}
K_1f_j=\lambda_j(\varphi)f_j,\qquad
\lambda_j(\varphi)=1+\rho\sqrt{\frac23}
 \cos\!\left(\varphi+\frac{2\pi j}{3}\right).
\label{eq:examples-k1-eigenvalues}
\end{equation}
All three directions remain fixed while their eigenvalues vary, so
$\r=3$. The index $j$ labels Fourier directions rather than physically
ordered charged masses. Assigning the three distinct positive mass
roots to those directions gives six discrete branches fixed by the charged
masses;
the charged columns are subsequently ordered as $e,\mu,\tau$.

\medskip
\noindent$\bullet$\;\textbf{$K_2$ ($\r=1$):}
The second pair first gives the identity-plus-traceless form
\begin{equation}
K_2=I+\frac{\rho\cos\theta}{\sqrt6}(3uu^T-I)
       +\frac{\rho\sin\theta}{\sqrt2}(uv^T+vu^T).
\label{eq:examples-k2-identity-form}
\end{equation}
Defining $\alpha=1-\rho\cos\theta/\sqrt6$ and
$\gamma=\rho\sin\theta/\sqrt2$, this can also be rewritten as
\begin{align}
K_2&=\alpha I+3(1-\alpha)uu^T+\gamma(uv^T+vu^T),
\label{eq:examples-k2-invariant}\\
&=\begin{pmatrix}
\alpha&0&2\beta\\
0&\alpha&-\beta\\
2\beta&-\beta&3-2\alpha
\end{pmatrix},\qquad \gamma=\sqrt5\,\beta.
\label{eq:examples-k2-matrix}
\end{align}
This reproduces the structure of Eq.~(14) in the currently available
preprint version (arXiv: 2610.10696v1) of Ref.~\cite{Strumia:2026aod},
where the corresponding texture was selected from a scan of a few
thousand similar possibilities. 
The coefficient matching and fixed-radius constraint are
\begin{equation}
\rho\cos\theta=\sqrt6(1-\alpha),\qquad
\rho\sin\theta=\sqrt2\,\gamma,\qquad
6(1-\alpha)^2+2\gamma^2=\rho^2.
\label{eq:examples-k2-coefficients}
\end{equation}
In particular, exact Koide requires
$6(1-\alpha)^2+10\beta^2=3$ and gives
$\cos\theta=\sqrt2(1-\alpha)$,
$\sin\theta=\sqrt{2/3}\,\gamma$. Thus $\alpha,\gamma$ are two
coefficients constrained to a circle, leaving one independent shape
parameter rather than two.

The direction perpendicular to both $u,v$ is
\begin{equation}
w=u\times v=(1,2,0)^T/\sqrt5,\qquad
T_2w=-\frac{w}{\sqrt6},\quad S_2w=0,\quad K_2w=\alpha w.
\label{eq:examples-k2-common-vector}
\end{equation}
This direction is fixed throughout the family, although its eigenvalue
$\alpha$ changes with the shape angle. There is no additional common
direction: $T_2$ has eigenvalue $2/\sqrt6$ along $u$ and eigenvalue
$-1/\sqrt6$ on $\operatorname{span}(v,w)$, while $S_2$ maps $u$ to
$v/\sqrt2$, $v$ to $u/\sqrt2$ and $w$ to zero. Only $w$ is common to
both generators. Hence $\r=1$; the remaining two directions generally
vary within the $(u,v)$ plane.

In the fixed orthonormal basis $Q=(u,v,w)$,
\begin{equation}
Q^TK_2Q=\begin{pmatrix}
3-2\alpha&\gamma&0\\
\gamma&\alpha&0\\
0&0&\alpha
\end{pmatrix}.
\label{eq:examples-k2-block}
\end{equation}
Its eigenvalues are
\begin{equation}
\lambda_w=\alpha,\qquad
\lambda_\pm=\frac{3-\alpha\pm
 \sqrt{9(1-\alpha)^2+4\gamma^2}}{2}.
\label{eq:examples-k2-eigenvalues}
\end{equation}
Matching the ordered charged roots fixes
\begin{equation}
\alpha=\lambda_\mu,\qquad
\gamma=\sqrt{(\lambda_\mu-\lambda_e)(\lambda_\tau-\lambda_\mu)}.
\label{eq:examples-k2-calibration}
\end{equation}
We take $\gamma>0$; the opposite sign is obtained by $v\mapsto-v$.
With $C,S_e>0$, the ordered charged diagonalizer is
\begin{equation}
U_e=(Cv-S_eu,\ w,\ S_ev+Cu),\qquad
C^2=\frac{\lambda_\tau-\lambda_\mu}{\lambda_\tau-\lambda_e},\quad
S_e^2=1-C^2.
\label{eq:examples-k2-diagonalizer}
\end{equation}
The notation $S_e$ distinguishes the mixing coefficient from the
matrix generator $S_2$. In particular, the common direction becomes
the muon eigenvector after matching to the charged masses.

\medskip
\noindent$\bullet$\;\textbf{$K_3$ ($\r=0$):}
We introduce the following explicit prescription to realize the third
category within the same framework:
\begin{equation}
K_3=I+aD+bH=
\begin{pmatrix}
1-a&b&2b\\
b&1&2b\\
2b&2b&1+a
\end{pmatrix}.
\label{eq:examples-k3-matrix}
\end{equation}
The real coefficients obey
\begin{equation}
\rho\cos\theta=\sqrt2\,a,\qquad
\rho\sin\theta=\sqrt{18}\,b,\qquad
2a^2+18b^2=\rho^2.
\label{eq:examples-k3-coefficients}
\end{equation}
Indeed, $\operatorname{tr}D^2=2$, $\operatorname{tr}H^2=18$ and
$\operatorname{tr}(DH)=0$, so
\begin{equation}
\operatorname{tr}K_3=3,\qquad
\operatorname{tr}K_3^2=3+2a^2+18b^2=3+\rho^2.
\label{eq:examples-k3-traces}
\end{equation}
Exact Koide therefore gives $a^2+9b^2=3/2$, with
$\cos\theta=\sqrt{2/3}\,a$ and $\sin\theta=\sqrt6\,b$.
Again only one coefficient angle is independent once the radius is fixed.

To establish $\r=0$, note that $D=\mathrm{diag}(-1, 0, 1)$ has three distinct eigenvalues, so
its only eigenvector directions are the coordinate axes. None is an
eigenvector of $H$, because each column of $H$ has a nonzero component
away from the corresponding axis. Thus $D,H$ have no common eigenvector.
No direction remains an eigenvector of the full $K_3(\theta)$ family;
this does not require each direction to move at every individual angle.
This $\r=0$ property also distinguishes the full generator family from
our $K_1,K_2$ families under any fixed unitary basis transformation. The explicit $D,H$ pattern is a phenomenological choice introduced here;
Koide's relation and the condition $\r=0$ do not uniquely select it.

The determinant supplies the remaining spectral information:
\begin{equation}
\det K_3=1-a^2-9b^2+3ab^2+8b^3
       =1-\frac{\rho^2}{2}+3ab^2+8b^3.
\label{eq:examples-k3-determinant}
\end{equation}
Its three eigenvalues are the roots of
\begin{equation}
\lambda^3 - \text{tr}(K_3)\lambda^2 + \mathrm{minor}(K_3)\lambda - \det(K_3) = 
\lambda^3-3\lambda^2+
 \left(3-\frac{\rho^2}{2}\right)\lambda-\det K_3=0.
\label{eq:examples-k3-characteristic}
\end{equation}
Here, `$\mathrm{minor}$' represents the sum of principal minors.
Matching the determinant in Eq.~\eqref{eq:examples-k3-determinant} to
$\lambda_e\lambda_\mu\lambda_\tau$, together with
Eq.~\eqref{eq:examples-k3-coefficients}, fixes the allowed shape angles.
For the supplied charged masses and the radius determined from them
below, there are two positive-spectrum branches, both retained in our
search. Their real ordered eigenvectors form an orthogonal diagonalizer
$O_3(\theta)$ satisfying
$O_3^TK_3O_3=\operatorname{diag}(\lambda_e,\lambda_\mu,\lambda_\tau)$.
Its directions depend on $\theta$ before matching to the charged masses.

\begin{figure}[b!]
\centering
\includegraphics[width=\linewidth]{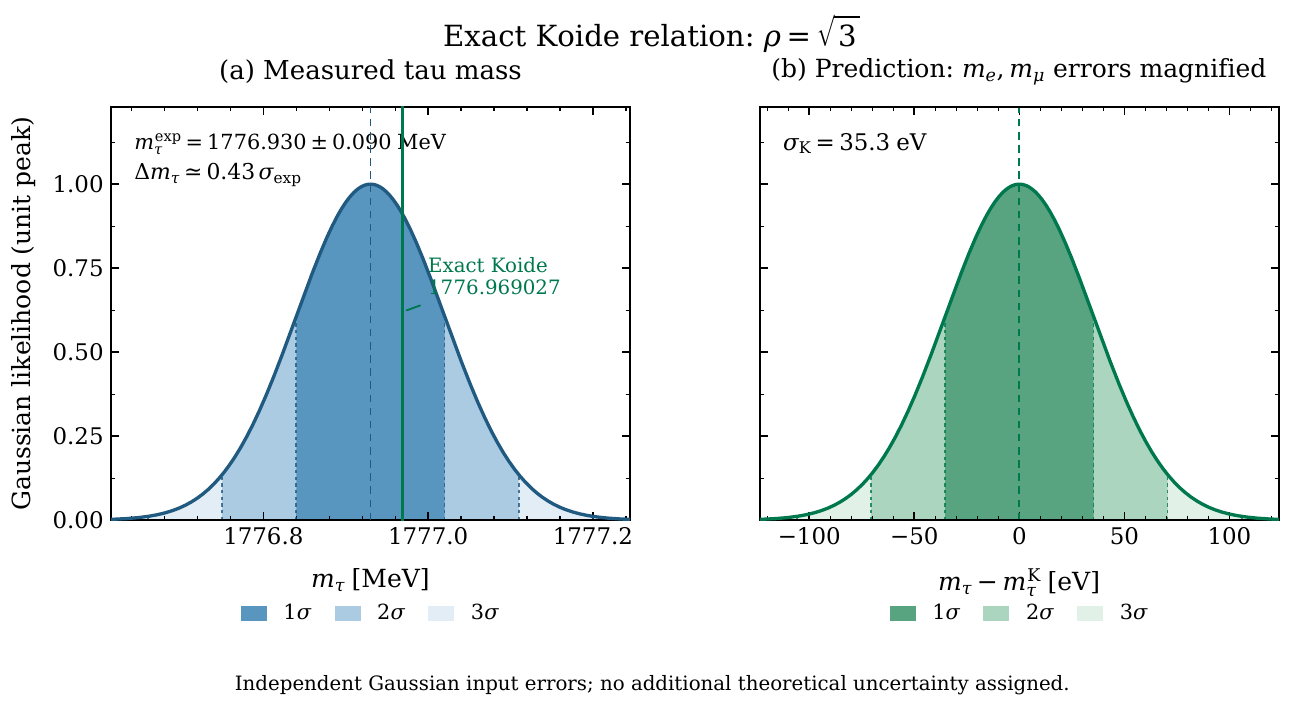}
\caption{Exact-Koide prediction for the tau mass using the electron and
muon inputs in Table~\ref{tab:experimental_inputs}.
Left: the measured tau mass is represented by a Gaussian with its
$1,2,3\sigma$ ranges; the green vertical line marks the Koide prediction.
The propagated prediction uncertainty is too narrow to resolve on this
scale. Right: the prediction is magnified on an eV offset scale,
showing $\sigma_{\rm K}\simeq35.3\,\mathrm{eV}$ and its $1,2,3\sigma$
ranges. Dark, medium and light shading denote the nested intervals.
The curves are individually normalized to unit peak. Input errors are
treated as independent and Gaussian; no additional theoretical
uncertainty is assigned to the exact relation.}
\label{fig:koide-tau-prediction}
\end{figure}

\subsection{Algebraic matching to charged-lepton masses and orientation}
\label{sec:charged-mass-matching}
Here matching means determining the parameters algebraically from the
supplied charged-lepton masses, rather than performing a statistical fit
or minimizing a goodness-of-fit function. For all three prescriptions,
positive $K_i$ with trace three requires
\begin{equation}
L=\sqrt{m_e}+\sqrt{m_\mu}+\sqrt{m_\tau},\qquad
\lambda_\ell=\frac{3\sqrt{m_\ell}}{L},\qquad
m_*=(L/3)^2.
\label{eq:examples-target-spectrum}
\end{equation}
Exact Koide sets $\rho=\sqrt3$. With $m_e,m_\mu$ fixed to their central
values, it predicts $m_\tau\simeq1.776969\,\mathrm{GeV}$, consistent with
the quoted $1.77693\pm0.00009\,\mathrm{GeV}$ at approximately $0.43\sigma$.
The explicit prediction is obtained from the heavy solution of
Eq.~\eqref{eq:koide}, with $m_\tau>m_\mu>m_e$:
\begin{equation}
m_\tau^{\rm K}=\left[2\bigl(\sqrt{m_e}+\sqrt{m_\mu}\bigr)
+\sqrt{3\bigl(m_e+m_\mu+4\sqrt{m_em_\mu}\bigr)}\right]^2.
\label{eq:koide-tau-mass-prediction}
\end{equation}
Treating the electron and muon mass errors in
Table~\ref{tab:experimental_inputs} as independent and Gaussian,
first-order propagation gives
\begin{equation}
\sigma_{\rm K}^2=\left(\frac{\partial m_\tau^{\rm K}}{\partial m_e}\right)^2
\sigma_e^2+\left(\frac{\partial m_\tau^{\rm K}}{\partial m_\mu}\right)^2
\sigma_\mu^2,\qquad
m_\tau^{\rm K}=(1776.969027\pm0.000035)\,\mathrm{MeV}.
\label{eq:koide-tau-mass-uncertainty}
\end{equation}
Here $\sigma_e,\sigma_\mu$ are the input $1\sigma$ errors, and
$\sigma_{\rm K}\simeq35.3\,\mathrm{eV}$ is dominated by the muon-mass
uncertainty. Figure~\ref{fig:koide-tau-prediction} compares this prediction
with the measured tau mass and displays the propagated $1,2,3\sigma$
ranges on a magnified scale. The predicted and measured central values differ by approximately \(0.43\sigma\), with the experimental and propagated prediction uncertainties combined in quadrature. The small prediction
uncertainty is conditional on exact Koide; it does not include an
additional theoretical uncertainty for the empirical relation.

From the above analysis it is clear that each viable texture pair reported in this work indeed admits a realization with \(\rho=\sqrt3\), satisfying the Koide relation exactly while reproducing all three charged-lepton masses within their experimental \(1\sigma\) uncertainties. In the following, we also consider an alternative
approach in which $\rho$ is adjusted to reproduce the experimental
central values of all three charged-lepton masses exactly. The catalog counts were obtained with central-mass matching; every retained texture pair was subsequently verified to admit an exact-Koide realization.

To match all three supplied central values exactly, the algebraic
matching instead fixes, without introducing an additional independent parameter,
\begin{equation}
Q_{\rm exp}=\frac{m_e+m_\mu+m_\tau}{L^2},\qquad
\rho=\sqrt{9Q_{\rm exp}-3}.
\label{eq:examples-measured-radius}
\end{equation}
The trace and trace-square then agree with the target spectrum, and
$\det K_i(\theta)=\lambda_e\lambda_\mu\lambda_\tau$ fixes its remaining
characteristic coefficient. The resulting discrete shape branches are
determined by the charged masses; they are not fitted to neutrino mixing.
Using the central masses in Table~\ref{tab:experimental_inputs}, the
common numerical inputs are
\begin{equation}
\begin{aligned}
(\lambda_e,\lambda_\mu,\lambda_\tau)
 &\simeq(0.040350442,\ 0.580216737,\ 2.379432821),\\
m_*&\simeq0.313850933\,\mathrm{GeV},\qquad
\rho\simeq1.732045083. ~~~\left( \sqrt{3}=1.7320508  \right)
\end{aligned}
\label{eq:charged-matching-numerical-inputs}
\end{equation}
Table~\ref{tab:charged-mass-matching} lists the resulting parameters
before any orientation is applied, with angles in $[0,2\pi)$. Effectively, we have a single shape parameter, namely \(\theta\), with \(\rho\simeq\sqrt3\) fixed and \(m_*\) setting the overall mass scale.
For $K_1$, the last column specifies the eigenvalues along the fixed
Fourier directions $(f_0,f_1,f_2)$; their six assignments give the six
shape-angle branches. The off-diagonal magnitude is common to all six,
$\rho/\sqrt6\simeq0.707104444$.
For $K_2$, the displayed branch has $\alpha=\lambda_\mu$ and
$\gamma=\sqrt5\,\beta\simeq0.985563860>0$; its opposite sign is already
covered by $v\mapsto-v$. For $K_3$, both mass-matched branches are shown.
All entries are rounded for presentation; the unrounded parameters
satisfy $m_*\lambda_\ell^2=m_\ell$ for all three charged leptons.

\begin{table}[t!]
\centering
\small
\renewcommand{\arraystretch}{1.15}
\begin{tabular}{cclc|c}
\toprule
Family & Branch & Parameter / Fourier assignment
 & Central-mass matching & Exact Koide\\
\midrule
\multicolumn{3}{r}{$\rho$} & $1.732045083342376$ & $\sqrt3$\\
\multicolumn{3}{r}{$m_*$ (GeV)} & $0.313850932981$ & $0.313856400291$\\
\multicolumn{3}{r}{$m_\tau$ (GeV)} & $1.776930000000$ & $1.776969027293$\\
\midrule
$K_1$ & 1 & $\theta;\ (\lambda_\tau,\lambda_e,\lambda_\mu)$
 & $0.222224762$ & $0.222222047$\\
 & 2 & $\theta;\ (\lambda_\mu,\lambda_e,\lambda_\tau)$
 & $1.872170340$ & $1.872173055$\\
 & 3 & $\theta;\ (\lambda_e,\lambda_\mu,\lambda_\tau)$
 & $2.316619864$ & $2.316617149$\\
 & 4 & $\theta;\ (\lambda_e,\lambda_\tau,\lambda_\mu)$
 & $3.966565443$ & $3.966568158$\\
 & 5 & $\theta;\ (\lambda_\mu,\lambda_\tau,\lambda_e)$
 & $4.411014967$ & $4.411012252$\\
 & 6 & $\theta;\ (\lambda_\tau,\lambda_\mu,\lambda_e)$
 & $6.060960545$ & $6.060963260$\\
\midrule
$K_2$ & 1 & $\theta$ & $0.935190493$ & $0.935184049$\\
 & & $\alpha=1-\frac{\rho\cos\theta}{\sqrt6}$ & $0.580216737$ & $0.580211684$\\
 & & $\beta=\frac{\rho\sin\theta}{\sqrt{10}}$ & $0.440757557$ & $0.440756919$\\
\midrule
$K_3$ & 1 & $\theta$ & $0.994866367$ & $0.994872626$\\
 & & $a=\frac{\rho\cos\theta}{\sqrt2}$ & $0.667012187$ & $0.667007962$\\
 & & $b=\frac{\rho\sin\theta}{\sqrt{18}}$ & $0.342391074$ & $0.342393597$\\
\cmidrule(lr){2-5}
 & 2 & $\theta$ & $1.551733725$ & $1.551727590$\\
 & & $a=\frac{\rho\cos\theta}{\sqrt2}$ & $0.023345333$ & $0.023352922$\\
 & & $b=\frac{\rho\sin\theta}{\sqrt{18}}$ & $0.408172769$ & $0.408174070$\\
\bottomrule
\end{tabular}
\caption{Parameters obtained by algebraic matching to the charged-lepton
central masses in Table~\ref{tab:experimental_inputs}, using the common
scale and radius in Eq.~\eqref{eq:charged-matching-numerical-inputs}.
For $K_1$, $\theta=\varphi$ and the listed triplets correspond to
$(f_0,f_1,f_2)$; for $K_2,K_3$, the coefficients refer to the explicit
matrices in Section~\ref{sec:koide-examples}. All angles are in radians. These values are fixed
before testing neutrino mixing.}
\label{tab:charged-mass-matching}
\end{table}

The six $K_1$ entries in Table~\ref{tab:charged-mass-matching} are
six parameter branches of the same circulant family, rather
than six independent texture families:
\begin{equation}
\theta=\pm\theta_0+\frac{2\pi n}{3}\pmod{2\pi},\qquad
\theta_0\simeq0.222224762,\qquad n=0,1,2.
\label{eq:k1-six-circulant-branches}
\end{equation}
The three branches for each sign cyclically permute the eigenvalue
assignments among the Fourier eigenvectors; changing the sign
complex-conjugates the matrix. All reproduce the same charged-lepton
masses. However, their physical equivalence must be checked
together with the neutrino matrix: a transformation of the charged
sector alone can change PMNS mixing. Table~\ref{tab:charged-mass-matching}
therefore lists six branches, without claiming six physically distinct
solutions.

For $K_3$, the two branches are the two solutions in $[0,2\pi)$ of
\begin{equation}
\det\!\left[I+\rho\bigl(\cos\theta\,T_3+\sin\theta\,S_3\bigr)\right]
=\lambda_e\lambda_\mu\lambda_\tau,
\label{eq:k3-mass-matching-branches}
\end{equation}
with the radius fixed by Eq.~\eqref{eq:examples-measured-radius}.
They occur at $\theta\simeq0.994866367$ and $1.551733725$, giving the
two $(a,b)$ pairs in Table~\ref{tab:charged-mass-matching}. Both matrices
have the same ordered charged-lepton eigenvalues but different
eigenvector directions. They are two mass-matched points of the same
fixed $(D,H)$ family. Both are retained until their physical equivalence
can be assessed together with the neutrino sector.

Once the charged masses have fixed these discrete branches, the
orientation of the charged eigenvectors relative to the neutrino sector
remains relevant to PMNS mixing. To explore this freedom, we apply a
fixed real orthogonal matrix $R$ to each mass-matched prescription:
\begin{equation}
K_i\mapsto RK_iR^T.
\label{eq:charged-family-orientation}
\end{equation}
This rotates both generators jointly, preserves the charged spectrum
and common-eigenvector count, and can change PMNS mixing relative to a
fixed neutrino matrix. The underlying generator pairs remain fixed
prescriptions; orientation scans alone do not enumerate every possible
pair in a category.

\section{Alikhanov Mass Relation}
\label{sec:alikhanov}
The neutrino mass spectrum has a milder hierarchy than the
charged-fermion spectra, while lepton mixing exhibits large angles.
In this work, we assume normal neutrino mass ordering,
$m_1<m_2<m_3$. The Alikhanov relation in Eq.~\eqref{eq:alikhanov}
can then be rewritten as
\begin{equation}
\frac{\sqrt{\Delta m_{31}^2}}{\sqrt{\Delta m_{21}^2}}
=3+2\sqrt2
\underset{m_1=0}{=}\frac{m_3}{m_2}. \label{eq:ak}
\end{equation}
The last equality holds only in the special case of a massless lightest
neutrino, for which
\begin{align}
m_1=0,\qquad m_2=\sqrt{\Delta m_{21}^2},\qquad
m_3=\sqrt{\Delta m_{31}^2}.
\end{align}
The simplest neutrino mass texture satisfying the Alikhanov
relation is given by~\cite{Strumia:2026aod}
\begin{equation}
M_\nu\propto
\begin{pmatrix}
0 & 0 & 0\\
0 & 0 & 1\\
0 & 1 & 2
\end{pmatrix}.
\end{equation}
Up to an overall proportionality constant, this matrix
yields the physical neutrino masses
\begin{equation}
m_1=0,\qquad
m_2=\sqrt{2}-1,\qquad
m_3=\sqrt{2}+1,
\end{equation}
which satisfy $m_3/m_2=3+2\sqrt{2}$, thereby reproducing the Alikhanov relation exactly.

\begin{figure}[t!]
\centering
\includegraphics[width=\linewidth]{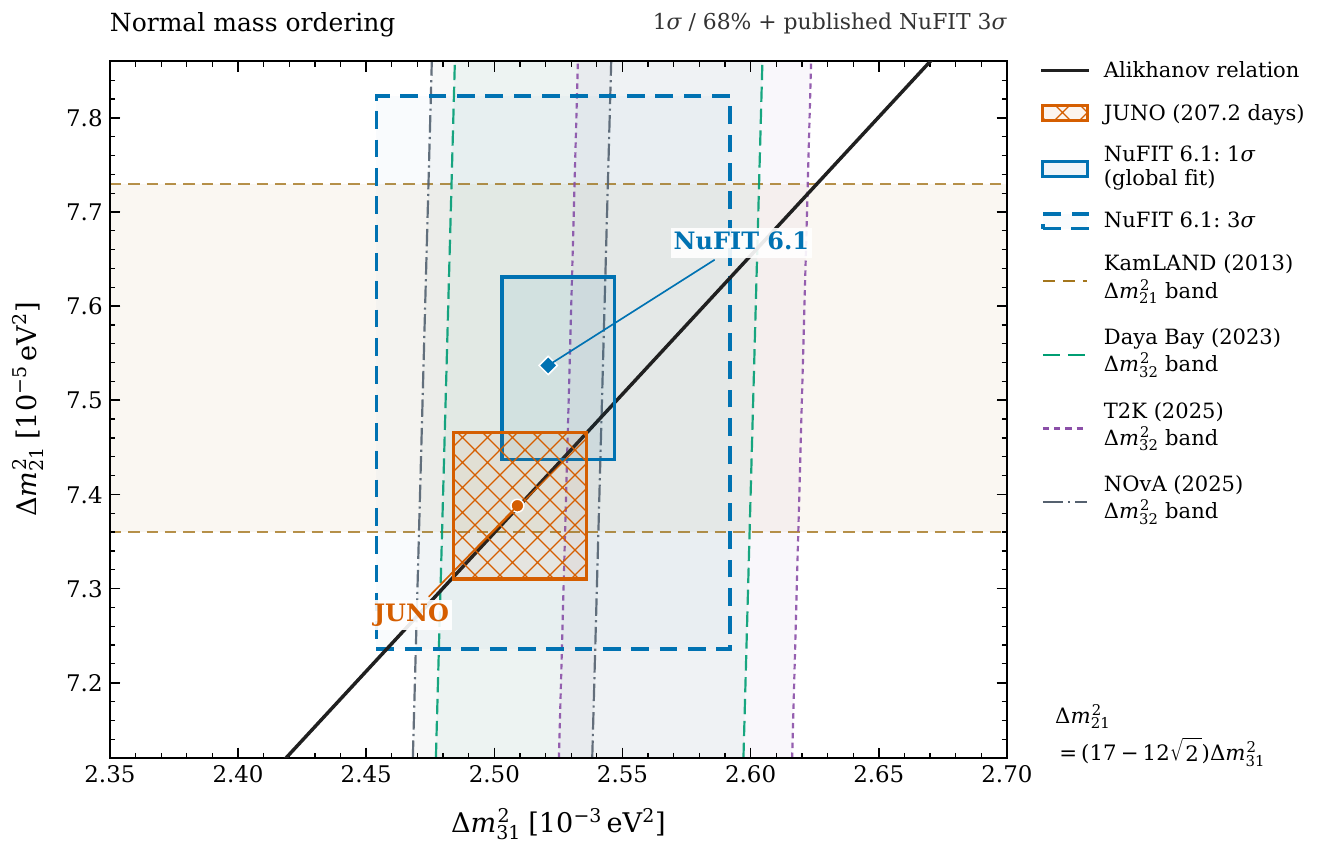}
\caption{
The black line is $\Delta m_{21}^2=(17-12\sqrt2)\Delta m_{31}^2$.
The orange box shows the JUNO $207.2$-live-day $1\sigma$ result~\cite{JUNO:2026tap};
the blue box with solid (dashed) edges shows NuFIT~6.1 $1\sigma$ ($3\sigma$) intervals~\cite{NUFIT}.
Markers
show the corresponding central values.
The horizontal band is the KamLAND-only $\Delta m_{21}^2$
interval~\cite{FigKamLAND2013}. The near-vertical bands show the
published marginal $\Delta m_{32}^2$ intervals from Daya Bay~\cite{FigDayaBay2023},
T2K~\cite{FigT2K2025} and NOvA~\cite{FigNOvA2025}, mapped using
$\Delta m_{31}^2=\Delta m_{32}^2+\Delta m_{21}^2$.
For NOvA we use its normal-ordering Bayesian result with the one-dimensional
reactor constraint on $\theta_{13}$.
Intervals retain their original $1\sigma$ confidence or $68\%$ credible
interpretation. The atmospheric bands illustrate marginal intervals,
without refitting the original likelihoods as the solar splitting varies.
The inputs overlap statistically: JUNO's atmospheric fit includes a Daya Bay
constraint, and NuFIT~\cite{NUFIT} is a global fit. The plot is a comparison of published
measurements, not a combined likelihood or future-sensitivity projection.}
\label{fig:alikhanov-splittings-comparison}
\end{figure}

For both $m_1=0$ and $m_1>0$, the Alikhanov relation is equivalently
\begin{equation}
q\equiv\frac{\Delta m_{21}^2}{\Delta m_{31}^2}
=17-12\sqrt2\simeq0.02943725,\qquad
m_2^2-m_1^2=q(m_3^2-m_1^2).
\label{eq:neutrino-splitting-ratio}
\end{equation}
We impose this relation exactly when selecting textures and subsequently
require both mass-squared differences to lie within the experimental
intervals in Table~\ref{tab:experimental_inputs}. Thus the relation need
not reproduce both experimental central values simultaneously.

Figure~\ref{fig:alikhanov-splittings-comparison} illustrates this relation
in the $(\Delta m_{31}^2,\Delta m_{21}^2)$ plane, extending the comparison
of Ref.~\cite{Alikhanov:2026bhi} to additional published measurements.
The JUNO~\cite{JUNO:2026tap} analysis based on $207.2$ live days reports, for normal
ordering~\cite{JUNO:2026tap},
\begin{equation}
\Delta m_{21}^2=(7.388\pm0.078)\times10^{-5}\,\mathrm{eV}^2,\qquad
\Delta m_{31}^2=\left(2.509^{+0.027}_{-0.025}\right)
\times10^{-3}\,\mathrm{eV}^2.
\label{eq:juno-splittings-comparison}
\end{equation}
Its central point lies close to the Alikhanov line, and both its
marginal-interval box and the NuFIT~6.1~\cite{NUFIT} box intersect that line.
The plot also shows the KamLAND~\cite{FigKamLAND2013} solar-splitting interval and the
atmospheric-splitting intervals from Daya Bay~\cite{FigDayaBay2023}, T2K~\cite{FigT2K2025} and NOvA~\cite{FigNOvA2025}.
These comparisons illustrate compatibility at the displayed precision;
they do not establish an exact relation or provide a combined
statistical significance. In particular, the boxes are products of
one-dimensional marginal intervals, and the displayed inputs overlap
statistically. Figure~\ref{fig:alikhanov-splittings-comparison} shows
measured constraints, rather than a projection of future sensitivity.
For consistency with the completed searches, the acceptance intervals
remain those of Table~\ref{tab:experimental_inputs}; the newer JUNO
measurement is included here as an additional comparison, without
recomputing the texture catalog.

\subsection{Neutrino texture prescription}
We consider three light Majorana neutrinos and write their reference
mass matrix as
\begin{equation}
M_\nu^{(0)}=\mu_\nu A,\qquad \mu_\nu>0,\qquad
A=A^T=\begin{pmatrix}
A_{11}&A_{12}&A_{13}\\
A_{12}&A_{22}&A_{23}\\
A_{13}&A_{23}&A_{33}
\end{pmatrix},\qquad A_{ij}\in\mathbb Z.
\label{eq:neutrino-integer-texture}
\end{equation}
Here $\mu_\nu$ fixes the neutrino mass scale, while the dimensionless
entries of $A$ specify the texture. Restricting the reference matrix to
real symmetric integer entries provides a simple, finite prescription
once an entry bound is chosen. It includes the real textures considered
in Ref.~\cite{Strumia:2026aod}, but the integer assumption and the
particular bound used below are choices of our search, rather than
consequences of the Alikhanov relation.  

The physical masses are the singular values, not the signed eigenvalues
of $A$. Define
\begin{equation}
(s_1,s_2,s_3)
=\operatorname{Sort}\!\left[\sqrt{\operatorname{Eigenvalues}(AA^\dagger)}\right],
\qquad m_i=\mu_\nu s_i,\qquad 0\le s_1<s_2<s_3.
\label{eq:neutrino-singular-values}
\end{equation}
For a real symmetric $A$, one may equivalently diagonalize
$V_\nu^TAV_\nu=\operatorname{diag}(\eta_1,\eta_2,\eta_3)$ and order the
columns so that $s_i=|\eta_i|$. Negative $\eta_i$ are retained: they do
not give negative physical masses. Instead, the Takagi diagonalizer is
\begin{equation}
W_\nu=V_\nu Q_\nu,\qquad
(Q_\nu)_{ii}=\begin{cases}1,&\eta_i\ge0,\\ i,&\eta_i<0,\end{cases}
\qquad W_\nu^TM_\nu^{(0)}W_\nu
=\operatorname{diag}(m_1,m_2,m_3).
\label{eq:neutrino-takagi}
\end{equation}
These column phases must be retained when computing Majorana-sensitive
observables such as neutrinoless double beta decay.

\subsection{Massless and massive lightest-neutrino cases}
The scale-independent selection condition is
\begin{equation}
\frac{s_2^2-s_1^2}{s_3^2-s_1^2}=q.
\label{eq:neutrino-texture-filter}
\end{equation}
We investigate both rank-two matrices, with $\det A=0$ and $s_1=0$,
and rank-three matrices, with $\det A\ne0$ and $s_1>0$. Rank-one and
zero matrices cannot provide the two observed nonzero splittings and
are excluded. In the rank-two case, Eq.~\eqref{eq:neutrino-texture-filter}
reduces to 
\begin{equation}
\frac{m_2}{m_3}=\frac{s_2}{s_3}=3-2\sqrt2.
\label{eq:neutrino-rank-two-ratio}
\end{equation}
For rank three, this fixed mass ratio is replaced by the splitting
condition in Eq.~\eqref{eq:neutrino-texture-filter}. The Alikhanov
relation itself leaves the absolute lightest mass undetermined.
Within our prescription, however, each integer matrix fixes all three
singular-value ratios, including $s_1/s_3$. Choosing an allowed
atmospheric splitting then fixes
\begin{equation}
\mu_\nu=\sqrt{\frac{\Delta m_{31}^2}{s_3^2-s_1^2}},\qquad
m_1=\sqrt{\Delta m_{31}^2}\,
\frac{s_1}{\sqrt{s_3^2-s_1^2}}.
\label{eq:neutrino-scale-matching}
\end{equation}
Thus $m_1$ is not an additional continuously adjustable parameter for
a fixed texture and scale. Its allowed range follows from the selected
matrix and the experimental splitting intervals.

Finally, satisfying the spectral relation does not determine whether
a texture gives acceptable lepton mixing. The flavor directions of
its eigenvectors must be compared with the charged-lepton eigenvectors
of Section~\ref{sec:koide}. We therefore retain relative flavor
placements and eigenvalue signs until the two sectors are combined.
The real reference textures are supplemented by the relative alignment
phases defined below; no experimental restriction on the neutrino
Dirac or Majorana phases is imposed.

\section{Results}
\label{sec:results}

\subsection{Numerical search strategy}
The search combines the charged-lepton prescriptions of
Section~\ref{sec:koide} with the neutrino matrices of
Section~\ref{sec:alikhanov}. The charged masses are fixed by algebraic
matching before any mixing test. The numerical task is then to identify
texture pairs and allowed relative alignments that satisfy the
oscillation intervals, rather than to fit all lepton masses and mixing
parameters simultaneously.

For the neutrino sector, we enumerate all six independent entries of
$A$ in $\{-5,-4,\ldots,5\}$. This gives
$11^6=1\,771\,561$ real symmetric matrices. The bound is imposed on the
dimensionless entries of $A$, not on the physical masses. Rank and exact
integer-invariant tests implement Eq.~\eqref{eq:neutrino-texture-filter};
independent singular-value calculations verify the retained spectra.
A common integer divisor is absorbed into $\mu_\nu$, and an overall
sign is redundant. Matrices are consequently reduced to primitive
representatives, whose six entries have greatest common divisor one.
Signed flavor permutations are used to organize the catalog, but
relative placements of the two sectors are retained when testing mixing.

For the charged sector, we retain all six mass-matched $K_1$ branches,
the $K_2$ prescription with $\alpha=\lambda_\mu$ and both signs covered
by the direction choices, and both $K_3$ branches in
Table~\ref{tab:charged-mass-matching}. The mass-matched coefficients
are not refitted to neutrino observables. We use a universal orientation prescription for all three
charged-lepton families. Introduce two nonzero primitive integer seed
vectors $\boldsymbol n_1,\boldsymbol n_2$ and construct
\begin{equation}
\begin{aligned}
\boldsymbol n_1\cdot\boldsymbol n_2&=0,\qquad
(n_k)_j\in\{-B,\ldots,B\},\quad k=1,2,\quad j=1,2,3,\\
\widehat{\boldsymbol n}_a&=\frac{\boldsymbol n_a}{\|\boldsymbol n_a\|},\qquad
E=\bigl(\widehat{\boldsymbol n}_1,\widehat{\boldsymbol n}_2,
\widehat{\boldsymbol n}_1\times\widehat{\boldsymbol n}_2\bigr).
\end{aligned}
\label{eq:search-integer-frame}
\end{equation}
Then $E^TE=I$ and $\det E=1$. The new symbols
$\boldsymbol n_k,\widehat{\boldsymbol n}_k$ describe the scanned frame;
the notation $u,v$ remains reserved for the fixed reference directions
of the earlier $K_2$ prescription. The integer bounds apply to the seed
components, while the normalized entries of $E$ generally contain square
roots.

To write the action of $E$ in the same form for every family, first
express each reference prescription in its own fixed orthonormal frame:
\begin{equation}
E_{1,0}=E_{3,0}=I,\qquad E_{2,0}=(u,v,w),\qquad w=u\times v.
\label{eq:search-reference-frames}
\end{equation}
Here $u,v$ are precisely those of Eq.~\eqref{eq:examples-k2-directions},
and $w$ is given in Eq.~\eqref{eq:examples-k2-common-vector}.
Define the reference-frame representations
\begin{equation}
\widetilde K_i=E_{i,0}^TK_iE_{i,0},\qquad
\widetilde T_i=E_{i,0}^TT_iE_{i,0},\qquad
\widetilde S_i=E_{i,0}^TS_iE_{i,0}.
\label{eq:search-reference-prescriptions}
\end{equation}
Thus $\widetilde K_1=K_1$ and $\widetilde K_3=K_3$, while
$\widetilde K_2$ is the block matrix already displayed in
Eq.~\eqref{eq:examples-k2-block}. These fixed reference-frame choices
introduce no additional free parameters.

Every candidate is now constructed by the same rule,
\begin{equation}
\begin{aligned}
K_i^{(E)}(\theta)&=E\widetilde K_i(\theta)E^T\\
&=I+\rho\left[\cos\theta\,(E\widetilde T_iE^T)
+\sin\theta\,(E\widetilde S_iE^T)\right],\qquad i=1,2,3.
\end{aligned}
\label{eq:search-universal-orientation}
\end{equation}
The two generators are rotated jointly, so the charged spectrum,
trace constraints and common-eigenvector category are preserved.
If $\widetilde U_{e,i}$ is the mass-ordered diagonalizer of the reference
prescription, the corresponding candidate has
$U_{e,i}^{(E)}=E\widetilde U_{e,i}$.
Relative to the original displayed matrices, the same construction is
$K_i^{(E)}=R_iK_iR_i^T$ with $R_i=EE_{i,0}^T$.
Consequently this uniform formulation reproduces the implemented
catalog exactly; it changes the notation and reference-frame
presentation, without changing the bounded candidates or reported
results.

For each candidate, the charged diagonalizer is ordered as
\begin{equation}
\begin{aligned}
U_e^\dagger M_eU_e&=\operatorname{diag}(m_e,m_\mu,m_\tau),\\
U_{\rm PMNS}=U&=U_e^\dagger\Phi W_\nu,\qquad
\Phi=\operatorname{diag}(e^{i\chi_1},e^{i\chi_2},1).
\end{aligned}
\label{eq:search-pmns}
\end{equation}
The columns of $W_\nu$ are likewise ordered as $(m_1,m_2,m_3)$. We retain the diagonal phase matrix $\Phi$, following the same approach as in Ref.~\cite{Strumia:2026aod}. 
Here $\chi_1,\chi_2$ are relative alignment phases between the two reference
textures; they are not the observable Dirac and Majorana phases. 
Equivalently, the neutrino matrix in the charged reference basis is
$\Phi^*M_\nu^{(0)}\Phi^\dagger$, whose Takagi diagonalizer is
$\Phi W_\nu$. For $B=2$, we allow the full two-phase domain. A separate
search with identity alignment, $\Phi=I$, extends the frame bound through
$B=5$. Thus free phases at larger frame bounds have not been exhaustively
scanned. The larger-bound identity cases can add solutions beyond the
smaller-bound free-phase catalog without contradicting the fact that
identity alignment is a special phase choice for a fixed texture pair.

Setting $\Phi=I$ removes the relative alignment phases, but does not
necessarily make both sectors real. The $K_1$ prescription is complex
Hermitian, so its charged-lepton diagonalization can generate a
nontrivial Dirac phase even for $\Phi=I$. In contrast, $K_2$, $K_3$ and
$A$ are real. With identity alignment these two charged prescriptions
conserve CP: $\delta_{\rm CP}=0$ or $\pi$, and the physical Majorana
phases are $0$ or $\pi$. For $m_1=0$, this statement applies to the
single physical Majorana-phase difference. A Majorana phase $\pi$ can
arise from a negative real neutrino eigenvalue: multiplying its
eigenvector by $i$ makes the Takagi mass positive and preserves CP.

The three mixing observables are extracted directly from $U$:
\begin{equation}
\sin^2\theta_{13}=|U_{e3}|^2,\qquad
\sin^2\theta_{12}=\frac{|U_{e2}|^2}{1-|U_{e3}|^2},\qquad
\sin^2\theta_{23}=\frac{|U_{\mu3}|^2}{1-|U_{e3}|^2}.
\label{eq:search-mixing-observables}
\end{equation}
A pair is accepted if all three lie simultaneously in the marginal
$3\sigma$ intervals of Table~\ref{tab:experimental_inputs}. This is an
interval-feasibility test, not a global likelihood fit. No experimental
cut is imposed on the Dirac or Majorana phases. 

The neutrino scale must also permit both splitting intervals. Imposing
Eq.~\eqref{eq:neutrino-splitting-ratio} gives the common allowed range
\begin{equation}
\Delta m_{31}^2\in
[2.4581099205,\,2.592]\times10^{-3}\,\mathrm{eV}^2.
\label{eq:search-splitting-scale}
\end{equation}
Saved benchmarks use $\Delta m_{31}^2=2.521\times10^{-3}\,\mathrm{eV}^2$;
then $\Delta m_{21}^2=q\Delta m_{31}^2$ is within its allowed $3\sigma$ interval.
Physical masses in both sectors are checked using ordered singular
values, and PMNS products are verified directly.

\subsection{Neutrino spectra selected by the exact relation}
Table~\ref{tab:neutrino-search-enumeration} summarizes the neutrino
catalog before any mixing requirement. Its counts are obtained by
exhaustive computational enumeration using exact algebraic filters.
Among the bounded rank-two
matrices, $240$ satisfy the exact relation; a further $144$ full-rank
matrices satisfy its nonzero-$m_1$ extension. Removing scale and overall
sign redundancies leaves $84$ and $72$ primitive matrices. These form
six signed-permutation classes in each rank. Such neutrino-only classes
are a way of organizing textures, not a count of physically distinct
combined lepton scenarios.

\begin{table}[t!]
\centering
\renewcommand{\arraystretch}{1.15}
\begin{tabular}{lrr}
\toprule
Selection & Rank two & Rank three\\
\midrule
Exact Alikhanov relation & $240$ & $144$\\
Primitive, with overall sign removed & $84$ & $72$\\
Placements after absorbing diagonal signs into free alignment & $27$ & $24$\\
Classes under signed flavor permutations & $6$ & $6$\\
\bottomrule
\end{tabular}
\caption{Neutrino enumeration within all $11^6$ real symmetric integer
matrices with independent entries in $[-5,5]$. There are $25\,786$
rank-two matrices before the exact-relation filter. Diagonal-sign
reduction is used for unrestricted alignment phases; the
identity-alignment search retains the corresponding relative signs.}
\label{tab:neutrino-search-enumeration}
\end{table}

The last row of Table~\ref{tab:neutrino-search-enumeration} means that
there are six neutrino-only texture classes for $m_1=0$ and six for
$m_1>0$. We can therefore give one explicit matrix for each class.
We denote these representatives by $\mathcal A_1,\ldots,\mathcal A_{12}$,
with labels following the numerical catalog; the calligraphic symbols
distinguish the class representatives from the individual entries
$A_{ij}$.\footnote{For example,
$\left(\begin{smallmatrix}2&1&0\\1&0&0\\0&0&0\end{smallmatrix}\right)$
and
$\left(\begin{smallmatrix}0&0&0\\0&0&-1\\0&-1&2\end{smallmatrix}\right)$
belong to the class represented by $\mathcal A_1$. The first follows
from simultaneously swapping its first and third rows and columns;
the second from changing the sign of the second flavor basis vector.
Both have the same ordered singular values as $\mathcal A_1$, hence
the same neutrino masses at fixed $\mu_\nu$. Their alignment with a
fixed charged-lepton matrix can nevertheless change PMNS mixing.
A common basis transformation of both sectors leaves physical mixing
unchanged.}

\medskip
\noindent\textbf{Rank two, $m_1=0$.} The six representatives are
\begin{equation}
\begin{aligned}
\mathcal A_{1}&=\begin{pmatrix}0&0&0\\0&0&1\\0&1&2\end{pmatrix},\qquad 
\mathcal A_{2}&=\begin{pmatrix}1&1&1\\1&1&1\\1&1&2\end{pmatrix},\qquad 
\mathcal A_{3}&=\begin{pmatrix}2&1&1\\1&5&5\\1&5&5\end{pmatrix},\\[1.2ex]
\mathcal A_{4}&=\begin{pmatrix}3&1&0\\1&1&0\\0&0&0\end{pmatrix},\qquad 
\mathcal A_{5}&=\begin{pmatrix}2&2&2\\2&1&1\\2&1&1\end{pmatrix},\qquad 
\mathcal A_{6}&=\begin{pmatrix}4&3&2\\3&2&2\\2&2&0\end{pmatrix}.
\end{aligned}
\label{eq:neutrino-six-rank-two-representatives}
\end{equation}

\medskip
\noindent\textbf{Rank three, $m_1>0$.} The six representatives are
\begin{equation}
\begin{aligned}
\mathcal A_{7}&=\begin{pmatrix}5&3&0\\3&-1&0\\0&0&2\end{pmatrix},\qquad 
\mathcal A_{8}&=\begin{pmatrix}5&3&0\\3&-1&0\\0&0&-2\end{pmatrix},\qquad 
\mathcal A_{9}&=\begin{pmatrix}4&2&1\\2&4&1\\1&1&-2\end{pmatrix},\\[1.2ex]
\mathcal A_{10}&=\begin{pmatrix}2&4&1\\4&2&1\\1&1&-2\end{pmatrix},\qquad 
\mathcal A_{11}&=\begin{pmatrix}2&3&3\\3&2&0\\3&0&2\end{pmatrix},\qquad 
\mathcal A_{12}&=\begin{pmatrix}2&3&3\\3&0&2\\3&2&0\end{pmatrix}.
\end{aligned}
\label{eq:neutrino-six-rank-three-representatives}
\end{equation}
Every retained bounded matrix belongs to the orbit of one of these
representatives:
\begin{equation}
A=\kappa\,\Pi^T\Sigma_\epsilon\mathcal A_\ell\Sigma_\epsilon\Pi,
\qquad \kappa\in\mathbb Z\setminus\{0\},\qquad
\Sigma_\epsilon=\operatorname{diag}(\epsilon_1,\epsilon_2,\epsilon_3),
\quad \epsilon_j=\pm1,
\label{eq:neutrino-representative-orbits}
\end{equation}
where $\Pi$ is a flavor-permutation matrix and only choices preserving
$|A_{ij}|\le5$ are included. The magnitude of $\kappa$ is absorbed into
the neutrino mass scale, and its sign is an unobservable overall sign.
These are classes under the specified real signed permutations and
scale identifications; their number is not a count of complete,
physically distinct lepton scenarios. Note that the textures $\mathcal A_{1,2,3}$ were previously considered in Ref.~\cite{Strumia:2026aod}.

All primitive full-rank textures satisfying the exact mass-relation filter have the same ordered singular-value spectrum, while the accepted rank-two textures share the mass ratio in Eq.~\eqref{eq:neutrino-rank-two-ratio}. At the benchmark atmospheric
splitting, the physical spectra are
\begin{equation}
\begin{aligned}
m_1=0:\quad &(m_1,m_2,m_3)
 \simeq(0,\ 0.008614599,\ 0.050209561)\,\mathrm{eV},\\
m_1>0:\quad &(m_1,m_2,m_3)
 \simeq(0.016981072,\ 0.019041222,\ 0.053003366)\,\mathrm{eV}.
\end{aligned}
\label{eq:search-benchmark-neutrino-spectra}
\end{equation}
Over the scale interval in Eq.~\eqref{eq:search-splitting-scale}, the
full-rank textures give $m_1\simeq0.016768$--$0.017219\,\mathrm{eV}$.
This narrow interval is a result of the bounded integer prescription;
it is not a general consequence of the Alikhanov relation.

The $6+6$ classes are the starting families for the mixing search,
but testing only the twelve displayed matrices in their displayed
flavor positions would be insufficient. A permutation of the neutrino
matrix alone changes its alignment relative to the charged-lepton
matrix and can change PMNS mixing. We therefore expand the classes
into all required relative flavor placements before pairing them with
the charged prescriptions. With unrestricted alignment phases, diagonal
sign changes can be absorbed into $\Phi$, leaving $27$ rank-two and
$24$ rank-three representatives to test. With identity alignment
$\Phi=I$, these relative signs must also be retained, so all $84$
rank-two and $72$ rank-three primitive matrices are tested.
Each is combined with the appropriate mass-matched charged branches
and bounded orientation frames described above. The mixing-angle cuts
are applied to these combined candidates, and their complete physical
prediction sets are compared subsequently.

\subsection{Combined textures and physical distinction}
Before imposing the mixing-angle cuts, the number of combinations is
larger than the nine reference charged branches: six for $K_1$, one for
$K_2$ and two for $K_3$. Each branch is supplemented by the bounded
orientation construction, and the resulting catalog sizes are given
in Table~\ref{tab:charged-search-catalog-sizes}.\footnote{For example, for $K_1$
at $B=2$, there are $98$ nonzero primitive integer vectors with entries
in $[-2,2]$. Their $1\,104$ ordered orthogonal pairs define frames
$R=(\widehat n_1,\widehat n_2,\widehat n_1\times\widehat n_2)$.
Combining each frame with all six charged-mass assignments produces
$1\,104\times6=6\,624$ raw parameterizations, representing $2\,040$
distinct charged matrices. For unrestricted alignment, ordered charged
diagonalizers related by $O'=D_fOD_\ell$, with diagonal unitary
$D_f,D_\ell$, are additionally identified: $D_f$ is absorbed into
$\Phi$, up to an irrelevant overall phase, while $D_\ell$ only
rephases charged mass eigenstates. This leaves $506$ free-alignment
catalog entries before any neutrino mixing test. The count is obtained
by explicit enumeration and removal of these redundancies, not by
counting $506$ viable charged--neutrino solutions.}

\begin{table}[th!]
\centering
\renewcommand{\arraystretch}{1.2}
\begin{tabular}{lrr}
\toprule
Charged prescription & Free alignment, $B=2$ & $\Phi=I$, $B\le5$\\
\midrule
$K_1$ & $506$ & $64\,824$\\
$K_2$ & $150$ & $12\,936$\\
$K_3$ & $552$ & $51\,744$\\
\midrule
Total charged entries & $1\,208$ & $129\,504$\\
Neutrino placements & $27+24=51$ & $84+72=156$\\
Pair parameterizations & $61\,608$ & $20\,202\,624$\\
\bottomrule
\end{tabular}
\caption{Catalog sizes before mixing-angle cuts, after redundancy
reductions. Charged counts include mass-matched branches and bounded
orientations; identity-alignment counts are cumulative through $B=5$.
Pair counts include the corresponding neutrino placements and count
parameterizations, not physically distinct scenarios.}
\label{tab:charged-search-catalog-sizes}
\end{table}

For unrestricted alignment phases, diagonal neutrino sign changes can
be absorbed into $\Phi$, so every charged entry is combined with
$27+24$ neutrino placements. Identity alignment retains the relative
signs and uses $84+72$ placements. Consequently, the two catalog
products give
\begin{equation}
\begin{aligned}
N_{\rm free}&=1\,208(27+24)=61\,608,\\
N_{\Phi=I}&=129\,504(84+72)=20\,202\,624,\\
N_{\rm catalog}&=N_{\rm free}+N_{\Phi=I}=20\,264\,232.
\end{aligned}
\label{eq:combined-search-catalog-size}
\end{equation}
Each free-alignment parameterization requires a feasibility test over
its continuous phase domain; its phase values are not counted as
separate discrete candidates. The two catalogs can overlap, and
common-basis redundancies and equivalent physical prediction families
are removed subsequently. Thus $N_{\rm catalog}$ is a bookkeeping
count of pair parameterizations, not the number of unique physical
possibilities or the number of numerical phase evaluations.

Combining the two alignment searches and removing structural duplicates
gives $403$ viable texture-pair labels. Each has a benchmark satisfying
the charged-lepton mass constraints and the individual $3\sigma$
intervals for both neutrino splittings and all three mixing angles.
Different labels may nevertheless describe the same physical predictions.

\begin{table}[t!]
\centering
\small
\renewcommand{\arraystretch}{1.15}
\setlength{\tabcolsep}{5pt}
\begin{tabular}{lc|rrrr|rrrr}
\toprule
 & & \multicolumn{4}{c}{Unrotated ($R_i=I$)}
 & \multicolumn{4}{c}{Bounded orientations}\\
\cmidrule(lr){3-6}\cmidrule(lr){7-10}
Neutrino texture & $m_1$ & $K_1$ & $K_2$ & $K_3$ & Total
 & $K_1$ & $K_2$ & $K_3$ & Total\\
\midrule
$\mathcal A_{1}$ & $0$ & 0 & 0 & 0 & 0 & 4 & 0 & 3 & 7\\
$\mathcal A_{2}$ & $0$ & 0 & 2 & 2 & 4 & 18 & 10 & 30 & 58\\
$\mathcal A_{3}$ & $0$ & 1 & 0 & 0 & 1 & 5 & 0 & 1 & 6\\
$\mathcal A_{4}$ & $0$ & 0 & 0 & 0 & 0 & 4 & 0 & 3 & 7\\
$\mathcal A_{5}$ & $0$ & 0 & 2 & 2 & 4 & 18 & 10 & 30 & 58\\
$\mathcal A_{6}$ & $0$ & 0 & 0 & 2 & 2 & 93 & 7 & 25 & 125\\
\midrule
Rank-two subtotal & $0$ & 1 & 4 & 6 & 11 & 142 & 27 & 92 & 261\\
\midrule
$\mathcal A_{7}$ & $>0$ & 0 & 0 & 0 & 0 & 4 & 0 & 3 & 7\\
$\mathcal A_{8}$ & $>0$ & 0 & 0 & 0 & 0 & 4 & 0 & $2^{\ast}$ & 6\\
$\mathcal A_{9}$ & $>0$ & 1 & 0 & 0 & 1 & 5 & 0 & 1 & 6\\
$\mathcal A_{10}$ & $>0$ & 1 & 0 & 0 & 1 & 5 & 0 & 1 & 6\\
$\mathcal A_{11}$ & $>0$ & 0 & 2 & 2 & 4 & 18 & 10 & 30 & 58\\
$\mathcal A_{12}$ & $>0$ & 0 & 2 & 2 & 4 & 18 & 10 & 30 & 58\\
\midrule
Full-rank subtotal & $>0$ & 2 & 4 & 4 & 10 & 54 & 20 & 67 & 141\\
\midrule
Total & & 3 & 8 & 10 & $\mathbf{21}$ & 196 & 47 & 159 & $\mathbf{402}$\\
\bottomrule
\end{tabular}
\caption{Viable physical prediction families for the unrotated
prescriptions and bounded orientations of
Eq.~\eqref{eq:charged-family-orientation}, at the stated tolerances.
The bounded counts include the $21$ unrotated families and combine
free alignment at $B=2$ with identity alignment through $B=5$.
${}^{\ast}$One $\mathcal A_8+K_3$ family duplicates an
$\mathcal A_7+K_3$ family and is counted only under $\mathcal A_7$.
Removing two contained $K_2$ families reduces $402$ to $400$
representatives without changing the prediction union.
Counts refer to families, not isolated points; all satisfy the
adopted absolute-mass bounds.}
\label{tab:combined-neutrino-family-counts}
\end{table}

For fixed ordered masses, two PMNS matrices describe the same physical
point when
\begin{equation}
U'=D_e U D_\nu,
\label{eq:search-physical-equivalence}
\end{equation}
where $D_e$ contains arbitrary charged-lepton phases. The diagonal entries
of $D_\nu$ are signs for massive Majorana states; the massless state,
when present, permits an arbitrary phase. Flavor labels remain fixed,
and CP-conjugate points are retained whenever physically inequivalent.
Any weak-basis transformation must act on both mass matrices.

Equality at one benchmark does not establish equivalence of entire
families. We therefore compare their full prediction sets, allowing
the complete alignment-phase domain for free-phase candidates and
keeping $\Phi=I$ for identity-alignment candidates.\footnote{The
comparison resolves $87\,882$ directed benchmark-to-family tests and
is validated on $1\,014$ independently generated self-family points.
}

One full-rank $K_3$ pair is analytically equivalent: its charged
diagonalizers coincide, while its neutrino matrices satisfy
$A'=J^TAJ$, with $J=\operatorname{diag}(1,1,i)$. The unrestricted
alignment phases absorb $J$. Removing this duplicate leaves
$402$ unequal physical prediction families at the stated numerical
tolerances.

Table~\ref{tab:combined-neutrino-family-counts} resolves the accepted
physical families by neutrino texture and charged-lepton prescription. In this table, for clarity, we have explicitly shown the unrotated (\(R_i=I\)) counts as well. This shows, allowing bounded rotations increases the count from 21 to 402, adding 381 unequal viable prediction families.  

Table~\ref{tab:combined-neutrino-family-counts} gives the resulting
$402$ unequal prediction families. Two rank-two $K_2$ families are
strictly contained in larger retained families; removing them leaves
$400$ representatives covering the same prediction union: $196$ from
$K_1$, $45$ from $K_2$ and $159$ from $K_3$. These counts refer to
families, not isolated parameter points. The bounded search finds
viable $K_1$ and $K_3$ partners for all twelve neutrino classes, but
does not exhaust all textures in the three common-eigenvector
categories or establish literature-wide novelty.

\subsection{Neutrino-mass predictions and observational constraints}
For each accepted scenario, we also compute the sum of neutrino
masses, the effective electron-neutrino mass, and the effective
Majorana mass relevant for neutrinoless double-beta decay,
respectively given by
\begin{equation}
\Sigma m_\nu=\sum_i m_i,\qquad
m_\beta=\sqrt{\sum_i|U_{ei}|^2m_i^2},\qquad
m_{\beta\beta}=\left|\sum_i U_{ei}^2m_i\right|.
\label{eq:search-effective-masses}
\end{equation}
We confront these predictions with the existing experimental
and cosmological constraints. For the sum of neutrino masses,
we adopt the cosmological upper bound
$\Sigma m_\nu<0.11\,\mathrm{eV}$~\cite{Naredo-Tuero:2024sgf,Gorbunov:2026sly}.
The laboratory constraints include the KATRIN bound on the
effective electron-neutrino mass,
$m_\beta<0.45\,\mathrm{eV}$ at $90\%$ confidence level~\cite{KATRIN:2024cdt},
and the neutrinoless double-beta decay limit from KamLAND-Zen,
$m_{\beta\beta}<L_{\beta\beta}$ at $90\%$ confidence level, where
$L_{\beta\beta}\in[28,122]\,\mathrm{meV}$
~\cite{KamLAND-Zen:2024eml}.
The latter interval reflects the uncertainties associated
with different nuclear matrix element calculations and
represents a range of upper limits rather than an allowed
interval for $m_{\beta\beta}$.
Since cosmological constraints on neutrino masses depend
on the underlying cosmological model and the datasets
employed, we do not impose the more stringent bound
$\Sigma m_\nu<0.072\,\mathrm{eV}$ obtained from analyses
incorporating DESI data~\cite{DESI:2024mwx}.

\begin{table}[t!]
\centering
\renewcommand{\arraystretch}{1.2}
\begin{tabular}{lccc}
\toprule
\multicolumn{4}{c}{Masses (eV)}\\
\midrule
Case & $m_1$ & $m_\beta$ & $m_{\beta\beta}$\\
\midrule
$m_1=0$ & $0$ & $0.008464$--$0.009336$ & $0.001293$--$0.004028$\\
$m_1>0$ & $0.016768$--$0.017219$ & $0.018783$--$0.019587$
 & $0.004071$--$0.018673$\\
\bottomrule
\end{tabular}
\par\medskip
\begin{tabular}{lcccc}
\toprule
\multicolumn{5}{c}{Phase benchmark spans (degrees)}\\
\midrule
Case & $\delta_{\rm CP}$ & $\alpha_{21}$ & $\alpha_{31}$
 & $\alpha_{31}-\alpha_{21}$\\
\midrule
$m_1=0$ & $0.00$--$359.80$ & --- & --- & $0.00$--$359.21$\\
$m_1>0$ & $0.00$--$359.77$ & $0.00$--$357.61$
 & $0.00$--$354.86$ & ---\\
\bottomrule
\end{tabular}
\caption{Mass and phase ranges for representative viable points of
the $261$ massless and $141$ massive-lightest-neutrino families.
The neutrino mass scale varies over
Eq.~\eqref{eq:search-splitting-scale}, with mixing parameters fixed.
Effective-mass and phase spans are not exhaustive allowed ranges.
All points satisfy $\Sigma m_\nu<0.11\,\mathrm{eV}$,
$m_\beta<0.45\,\mathrm{eV}$ and even the strongest adopted
$m_{\beta\beta}<28\,\mathrm{meV}$ limit.
Phases are in degrees modulo $360^\circ$; dashes denote unphysical
or redundant phases. }
\label{tab:viable-benchmark-mass-ranges}
\end{table}

Table~\ref{tab:viable-benchmark-mass-ranges} summarizes the neutrino-mass
predictions for representative viable points of the $402$ texture
families, grouped according to whether $m_1=0$ or $m_1>0$. At each
point, the mixing angles and phases are fixed, while the overall
neutrino mass scale varies over the interval in
Eq.~\eqref{eq:search-splitting-scale}. The masses $m_i$, $m_\beta$, and $m_{\beta\beta}$ consequently
scale as $\sqrt{\Delta m_{31}^2}$.
The quoted ranges of the effective masses $m_\beta$ and
$m_{\beta\beta}$ characterize these representative points and
need not exhaust the allowed ranges of each family. This is because the alignment phases $\chi_1,\chi_2$ are constrained,
but generally not uniquely fixed, by requiring that the three neutrino
mixing angles to lie within their experimental $3\sigma$ intervals. In contrast, the common full-rank spectrum and the allowed
mass-splitting interval determine the full range of $m_1$,
independently of the alignment phases.

We also report the physical phases for which we use the standard PMNS convention
\begin{equation}
U=V(\theta_{12},\theta_{23},\theta_{13},\delta_{\rm CP})
\operatorname{diag}\!\left(1,e^{i\alpha_{21}/2},e^{i\alpha_{31}/2}\right).
\label{eq:search-phase-convention}
\end{equation}
For $m_1>0$, both Majorana phases are physical. For $m_1=0$, only
$\alpha_{31}-\alpha_{21}$ is independent of the arbitrary phase of the
massless state. Phases are defined modulo $2\pi$ and tabulated in degrees.
These output values are obtained without experimental phase cuts.

\begin{figure}[t!]
\centering
\includegraphics[width=\linewidth]{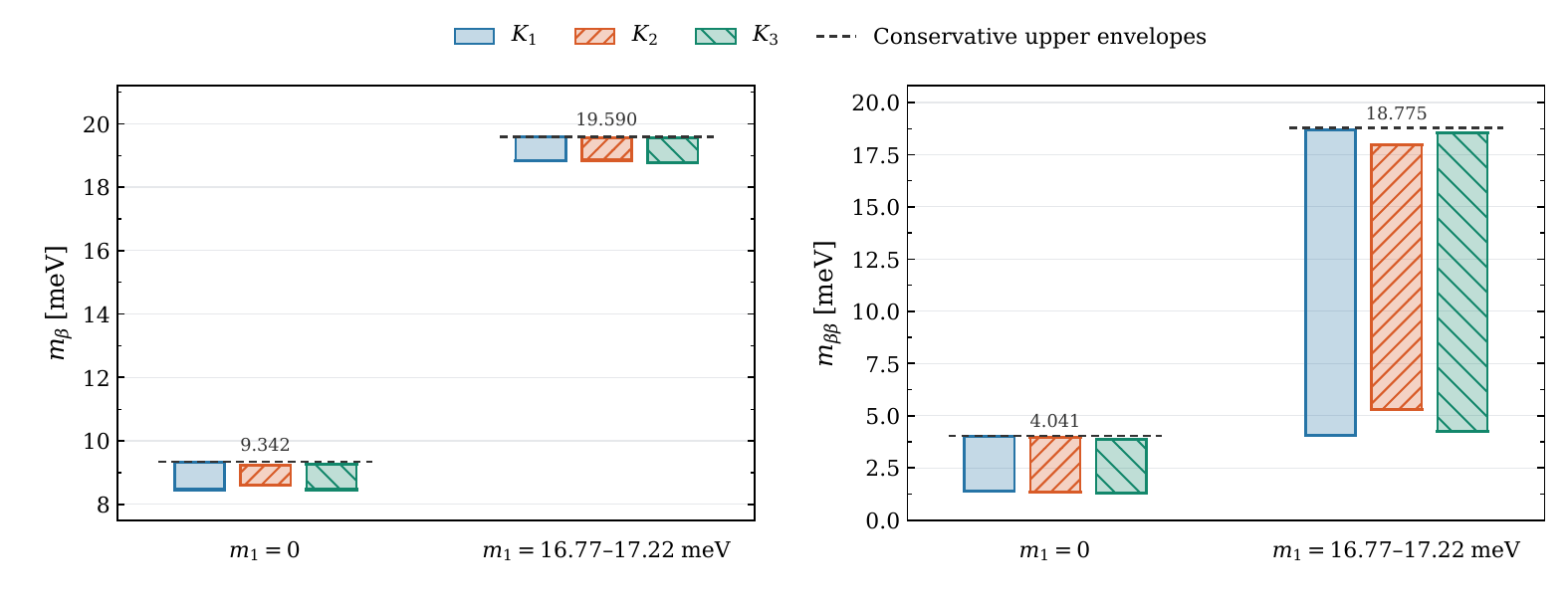}
\caption{Representative ranges of $m_\beta$ (left) and
$m_{\beta\beta}$ (right) for $K_1,K_2,K_3$ in the two neutrino-mass
branches, including the allowed mass-scale variation.
Dashed lines indicate conservative upper bounds.
The bars do not exhaust the continuous allowed regions.}
\label{fig:effective}
\end{figure}

To check the continuous allowed regions,  we use the common scale interval in
Eq.~\eqref{eq:search-splitting-scale}, the full allowed mixing-angle
intervals and the phase-independent inequality
\begin{equation}
m_{\beta\beta}\le\sum_i |U_{ei}|^2m_i.
\label{eq:search-effective-mass-envelope}
\end{equation}
The conservative bounds in Table~\ref{tab:search-absolute-mass-bounds}
apply to every retained texture and every allowed phase choice (they
need not be attained simultaneously by any single texture). Moreover, Figure~\ref{fig:effective} compares the representative
effective-mass ranges for the three charged-lepton prescriptions
in the two neutrino-mass branches.

\begin{table}[th!]
\centering
\renewcommand{\arraystretch}{1.2}
\begin{tabular}{lrrr}
\toprule
Spectrum & $\Sigma m_\nu$ & $m_\beta$ & $m_{\beta\beta}$\\
\midrule
$m_1=0$ & $\le59.647$ & $\le9.342$ & $\le4.041$\\
$m_1>0$ & $\le90.271$ & $\le19.590$ & $\le18.775$\\
\bottomrule
\end{tabular}
\caption{Conservative upper envelopes in meV, rounded upward, over the
allowed scale, mixing-angle intervals and unrestricted neutrino phases.}
\label{tab:search-absolute-mass-bounds}
\end{table}

\subsection{Explicit benchmarks for the \texorpdfstring{$\mathcal A_1$}{A1} class}
\label{sec:a1-benchmarks}
For completeness, we present representative viable solutions
from the $\mathcal A_1$ class. These seven benchmark examples ($\mathcal B_1,\ldots,\mathcal B_7$) illustrate how the
textures and their predictions can be reconstructed explicitly.

Table~\ref{tab:a1-benchmark-frames} specifies one viable benchmark from
each of the seven $\mathcal A_1$ families in
Table~\ref{tab:combined-neutrino-family-counts}. Together with the fixed
$K_1,K_3$ generators in Eq.~\eqref{eq:examples-fixed-matrices} and
Table~\ref{tab:generator-angle-table}, these inputs fully determine both
mass matrices:
\begin{equation}
\begin{aligned}
R_s&=\bigl(\widehat{\boldsymbol n}_1^{(s)},\,
\widehat{\boldsymbol n}_2^{(s)},\,
\widehat{\boldsymbol n}_1^{(s)}\times\widehat{\boldsymbol n}_2^{(s)}\bigr),\\
K^{(s)}&=R_sK_{i(s)}(\theta_s)R_s^T,\qquad
K_{i(s)}(\theta_s)=I+\rho\bigl(\cos\theta_s\,T_{i(s)}
+\sin\theta_s\,S_{i(s)}\bigr),\\
M_e^{(s)}&=m_*[K^{(s)}]^2,\qquad
M_\nu^{(s)}=\Phi_s^*(\mu_\nu\mathcal A_1)\Phi_s^\dagger.
\end{aligned}
\label{eq:a1-benchmark-matrices}
\end{equation}
When forming PMNS mixing, the charged masses
and their eigenvector columns are sorted together as $(e,\mu,\tau)$,
and the neutrino masses and their Takagi columns as $(m_1,m_2,m_3)$.

\begin{table}[b!]
\centering
\small
\renewcommand{\arraystretch}{1.4}
\begin{tabular}{ccc}
\toprule
\multicolumn{3}{c}{Common inputs for all seven benchmarks}\\
\midrule
$\rho$ & $m_*$ (GeV) & $\mu_\nu$ (eV)\\
1.732045083342376 & 0.3138509329807888 & 0.02079748106177416\\
\bottomrule
\end{tabular}
\par\medskip
\renewcommand{\arraystretch}{1.9}
\setlength{\tabcolsep}{4pt}
\begin{tabular}{lccccc}
\toprule
Case & Charged & $\theta_s$ (rad) & $\widehat{\boldsymbol n}_1^{(s)}$
 & $\widehat{\boldsymbol n}_2^{(s)}$ & $\Phi_s$\\
\midrule
$\mathcal B_1$ & $K_1$ & 2.316619864297221 & $\dfrac{(0,3,-2)^T}{\sqrt{13}}$ & $\dfrac{(4,2,3)^T}{\sqrt{29}}$ & $I$\\
$\mathcal B_2$ & $K_1$ & 3.966565442882365 & $\dfrac{(0,3,-2)^T}{\sqrt{13}}$ & $\dfrac{(4,2,3)^T}{\sqrt{29}}$ & $I$\\
$\mathcal B_3$ & $K_1$ & 2.316619864297221 & $\dfrac{(5,3,5)^T}{\sqrt{59}}$ & $\dfrac{(0,-5,3)^T}{\sqrt{34}}$ & $I$\\
$\mathcal B_4$ & $K_1$ & 3.966565442882365 & $\dfrac{(5,3,5)^T}{\sqrt{59}}$ & $\dfrac{(0,-5,3)^T}{\sqrt{34}}$ & $I$\\
$\mathcal B_5$ & $K_3$ & 0.994866367389440 & $\dfrac{(-1,-1,-1)^T}{\sqrt{3}}$ & $\dfrac{(0,1,-1)^T}{\sqrt{2}}$ & $\operatorname{diag}(-1,e^{11\pi i/12},1)$\\
$\mathcal B_6$ & $K_3$ & 0.994866367389440 & $\dfrac{(-2,3,0)^T}{\sqrt{13}}$ & $(0,0,1)^T$ & $I$\\
$\mathcal B_7$ & $K_3$ & 0.994866367389440 & $\dfrac{(-5,4,-3)^T}{\sqrt{50}}$ & $\dfrac{(4,5,0)^T}{\sqrt{41}}$ & $I$\\
\bottomrule
\end{tabular}
\caption{Complete reconstruction inputs for the seven $\mathcal A_1$
benchmarks (see Table~\ref{tab:combined-neutrino-family-counts}), using the unrotated $K_1,K_3$ conventions of Section~2.3.
All three $K_3$ entries use the first mass-matched branch.}
\label{tab:a1-benchmark-frames}
\end{table}

\begin{table}[t!]
\centering
\small
\setlength{\tabcolsep}{4pt}
\renewcommand{\arraystretch}{1.25}
\begin{tabular}{lrrrrrrr}
\toprule
Case & $\sin^2\theta_{12}$ & $\sin^2\theta_{23}$ & $\sin^2\theta_{13}$
 & $\delta_{\rm CP}$ & $\alpha_{31}-\alpha_{21}$
 & $m_\beta$ & $m_{\beta\beta}$\\
\midrule
$\mathcal B_1$ & 0.31907157 & 0.50000000 & 0.02357098 & 90.00000 & 0.00000 & 9.08531991 & 1.50039587\\
$\mathcal B_2$ & 0.31907157 & 0.50000000 & 0.02357098 & 270.00000 & 0.00000 & 9.08531991 & 1.50039587\\
$\mathcal B_3$ & 0.32264626 & 0.50000000 & 0.02153689 & 90.00000 & 0.00000 & 8.81605519 & 1.63824904\\
$\mathcal B_4$ & 0.32264626 & 0.50000000 & 0.02153689 & 270.00000 & 0.00000 & 8.81605519 & 1.63824904\\
$\mathcal B_5$ & 0.29301629 & 0.44385453 & 0.02370312 & 160.87262 & 197.70521 & 8.99918161 & 2.05086675\\
$\mathcal B_6$ & 0.31983581 & 0.45575849 & 0.02263611 & 180.00000 & 180.00000 & 8.95900574 & 1.55633983\\
$\mathcal B_7$ & 0.30554479 & 0.54241289 & 0.02311339 & 0.00000 & 180.00000 & 8.96769988 & 1.41079476\\
\bottomrule
\end{tabular}
\caption{Mixing and effective-mass predictions for the seven
$\mathcal A_1$ benchmarks. Phases are in degrees in the convention of
Eq.~\eqref{eq:search-phase-convention}; effective masses are in meV.
Since $m_1=0$, only one independent Majorana-phase combination is
reported. All points satisfy the adopted $3\sigma$ mixing intervals
and absolute-mass bounds.}
\label{tab:a1-benchmark-mixing}
\end{table}

The neutrino mixing parameters and effective-mass predictions for these seven
$\mathcal A_1$ benchmarks are listed in Table~\ref{tab:a1-benchmark-mixing}.
The pairs $(\mathcal B_1,\mathcal B_2)$ and $(\mathcal B_3,\mathcal B_4)$ are CP conjugates, retained
separately in our count. The real identity-alignment points $\mathcal B_6,\mathcal B_7$
conserve CP. Only $\mathcal B_5$ uses nontrivial alignment phases; it illustrates
one point of a continuous accepted family, not its full prediction range.

\section{Conclusions}
\label{sec:conclusions}
We have systematically investigated charged-lepton and neutrino
mass-matrix textures motivated by the empirical Koide and Alikhanov
relations, confronting their predictions for lepton mixing with
experimental data.  For the charged-lepton sector, we classify positive Hermitian
square-root matrix families into \textbf{three distinct} categories according to
the number of common eigenvector directions of their generators:
three, one, or none. We investigate one explicit family from
each category.

An exhaustive search of real symmetric integer neutrino matrices
with entries between $-5$ and $5$ identifies \textbf{twelve distinct}
signed-permutation classes, six with $m_1=0$ and six with $m_1>0$.
Combining these with the charged-lepton constructions yields
$403$ viable structural pairs, corresponding to $402$ distinct
physical prediction families. The complete set of
allowed predictions can be reproduced using  \textbf{400}
representative families. All twelve neutrino classes admit
phenomenologically viable charged-lepton partners.

Our analysis identifies two distinct neutrino mass branches:
a massless-lightest-neutrino scenario ($m_1=0$) and a full-rank
scenario predicting $m_1\simeq16.77$--$17.22\,\mathrm{meV}$. Representative viable solutions yield approximately \(m_{\beta\beta}\simeq1.29\)–\(4.03\) meV and \(4.07\)–\(18.67\) meV for the two branches, respectively. More generally, all viable solutions satisfy
the conservative upper bounds $m_{\beta\beta}\le4.041\,\mathrm{meV}$
for $m_1=0$ and $m_{\beta\beta}\le18.775\,\mathrm{meV}$
for $m_1>0$. All viable solutions satisfy the adopted
cosmological ($\Sigma m_\nu < 0.11\,\mathrm{eV}$) and laboratory neutrino-mass constraints,
although stronger cosmological bound $\Sigma m_\nu < 0.072\,\mathrm{eV}$ from  the DESI collaboration~\cite{DESI:2024mwx} can exclude the
full-rank branch.

While our classification is restricted to the specified
search domain and charged-lepton prescriptions, it
demonstrates that a remarkably rich variety of lepton mass
textures can accommodate both empirical relations while
reproducing the observed mixing angles. Future improvements
in neutrino mass-squared difference measurements, particularly
from JUNO, will further test the Alikhanov relation, while
absolute neutrino mass measurements offer a means of
distinguishing the two spectral branches.

\section*{Acknowledgments}
The numerical scans presented in Sec.~4 were facilitated
with the assistance of OpenAI's GPT-6.1-Sol model. S.S.\  acknowledges the financial support from the Slovenian Research Agency (research core funding No.\ P1-0035). 
\bibliographystyle{style}
\bibliography{reference}
\end{document}